\documentclass[a4paper, amsfonts, amssymb, amsmath, reprint, showkeys, nofootinbib, twoside, floatfix]{revtex4-1}
\usepackage[english]{babel}
\usepackage[utf8]{inputenc}
\usepackage[colorinlistoftodos, color=green!40, prependcaption]{todonotes}
\usepackage{amssymb}
\usepackage[export]{adjustbox}
\usepackage{amsthm}
\usepackage{mathtools}
\usepackage{physics}
\usepackage{xcolor}
\usepackage{graphicx}
\usepackage[left=23mm,right=13mm,top=35mm,columnsep=15pt]{geometry} 
\usepackage{adjustbox}
\usepackage{placeins}
\usepackage[T1]{fontenc}
\usepackage{lipsum}
\usepackage{csquotes}
\usepackage[pdftex, pdftitle={Article}, pdfauthor={Author}]{hyperref} 
\begin{document}
\title{An Algorithm for Fast Supervised Learning in Variational Circuits through Simultaneous Processing of Multiple Samples}

\author{Siddharth Dangwal, Ritvik Sharma, and Debanjan Bhowmik}
    \email[Correspondence email address: ]{debanjan@ee.iitd.ac.in}
    \affiliation{Department of Electrical Engineering, Indian Institute of Technology Delhi, New Delhi 110016, India}


\begin{abstract}
We propose a novel algorithm for fast training of variational classifiers by processing multiple samples parallelly. The algorithm can be adapted for any ansatz used in the variational circuit. The presented algorithm utilizes qRAM and other quantum circuits in the forward pass. Further, instead of the usual practice of computing the loss classically, we calculate the loss using a Swap-test circuit. The algorithm thus brings down  the training cost of a variational classifier to $\mathcal{O}(\log N)$ from the usual $\mathcal{O}(N)$ when training on a dataset of $N$ samples. Although we discuss only binary classification in the paper, the algorithm can be easily generalized to multi-class classification.

\end{abstract}

\maketitle

\section{Introduction} \label{sec:introduction}
In recent years, Machine Learning (ML) has become a fascinating area of research due to its successful application to problems like image and speech recognition, language processing, predicting health hazards and natural catastrophes, material science, etc. \cite{lecun2015,mikolov2013distributed, schmidt2019recent, wernick2010machine, 726791} This success arises from its fairly unique data-driven approach, instead of hard-coding rules and instructions to the computer, for carrying out classification and object detection tasks. In ML, we supply data to the computer, and the computer attempts to solve an optimization problem to learn the required instructions. \cite{lecun2015}.

Quantum computing has turned out to be another fascinating area for research in recent years. Using counter-intuitive phenomena like superposition and entanglement, which are not experienced in the macroscopic classical world but are regularly experienced in the microscopic quantum world, an advantage in terms of speed of computation has been shown for various quantum algorithms compared to their classical counterparts \cite{Preskill2018,AlainAspectIntro,NielsenChuang,google2018}.

Recently, various quantum algorithms have been proposed as alternatives to popular, classical algorithms for carrying out ML tasks. \cite{MariaNatureCommentary2019,IJCNNQMLReview2020,NASAreview,biamonte2017quantum, schuld2019quantum, adhikary2020}. ML often involves manipulating high-dimensional vectors for data classification. Quantum algorithms utilize quantum parallelism through superposition and entanglement and are considered useful for handling such high-dimensional vectors for ML tasks \cite{biamonte2017quantum, schuld2019quantum,lloyd2013quantum}. This insight has led to the development of these Quantum Machine Learning (QML) algorithms \cite{lloyd2013quantum,rebentrost2014quantum,lloyd2018quantum, cong2019quantum,farhi2014quantum, peruzzo2014variational}.

Variational quantum algorithms, which are essentially hybrid quantum-classical algorithms using a parameterized quantum circuit, form an important subclass of QML algorithms. In variational algorithms, as shown in Fig. \ref{fig:ansatz}, the values of the parameters in the feedforward quantum circuit, also known as the quantum ansatz, are updated every epoch through a classical feedback loop to reach a desired point in the parameter space after several epochs. At this point, the value of a pre-defined loss function, dependent on these parameters and the samples in a dataset (on which classification needs to be achieved), attains a minimum value \cite{schuld2020circuit, farhi2014quantum, peruzzo2014variational,markovic2020quantum,schuld2019transferlearning}. Thus supervised learning is achieved in the variational quantum circuit corresponding to that dataset.

In such a variational algorithm, the total loss calculation for all the samples per epoch is carried out by passing forward one sample at a time through the quantum ansatz (more details in Section II below) \cite{havlivcek2019, Tachino, grant2018hierarchical}. On the contrary, the loss optimization through updating of parameters in the quantum ansatz (classical feedback loop) is carried out only a few times per epoch (batch training) \cite{bishop2006pattern,haykin2010neural,goodfellow2016deep}. Thus, the forward passing \text{on the forward pass of all samples} of all the samples and the corresponding total loss calculation steps form the major bottleneck in the learning process. Hence, the time complexity of the algorithm, as dominated by these steps as opposed to the loss optimization step, can be expressed in terms of the number of input samples as $\mathcal{O}(N)$ where the number of samples in the training set is $N$. 

In this context, this paper proposes a novel algorithm for training any arbitrary parameterized ansatz (with the constraint that the data encoding scheme used be non-parameterized), where the time complexity of the algorithm, again dominated by the forward pass and the loss calculation steps, is $\mathcal{O}(\log N)$. The algorithm is implemented using circuits that encode multiple data points into a single quantum state and apply parameterized operations on that state to implement the forward pass. The loss calculation is also done using a quantum circuit instead of the conventional case, where it is computed classically, an idea explored in Ref. \cite{cao2020cost}. The loss is then differentiated classically as in general variational algorithms, and the parameters are updated till they correspond to the loss function's minima. We discuss these steps in detail in Section \ref{sec:algorithm} and also describe the quantum circuits implementing them. Before the first epoch of the training process, we encode all the samples from the classical data set (for training) to corresponding quantum states and store them in a qRAM \cite{giovannetti2008quantum}. As a result, each sample does not need to be encoded into a quantum state in every subsequent epoch. Instead, the quantum states can be retrieved from the qRAM at the beginning of every epoch for the processing. We retrieve multiple samples for processing from the qRAM as a superposition of samples correlated with their addresses. All subsequent quantum operations act on this superposition of different sample states correlated with their addresses. Thus, doing this gives us a computational cost advantage.  This computational advantage would be absent if we processed a single sample at a time, as explained in section ~\ref{sec:algorithm}.

Thus, our proposed algorithm 
provides an exponential speed-up for the forward pass, which would 
ameliorate supervised learning performance in quantum computing systems. To the best of our knowledge, most QML algorithms focus on achieving complexity advantage with respect to the input dimensions and not the number of input samples \cite{schuld2019quantum, schuld2020circuit, giovannetti2008quantum}. On the contrary, we focus on bringing about a complexity advantage in terms of the number of input samples. 

In Section \ref{sec:background}, we first provide the basic background for our variational algorithm . Then, in section \ref{sec:algorithm} we discuss the various steps in our proposed algorithm. In Section \ref{sec:experiments}, we use our algorithm to achieve supervised learning on Fisher's Iris data set, on which we report high classification accuracy numbers. Next, in section \ref{sec:discussion}, we discuss the computational complexity and then the advantages and limitations of our proposed algorithm. In Section V, we conclude the paper.

\section{Background} \label{sec:background}

\subsection{The Classification Task}\label{cl_task}

First, we briefly describe the binary classification problem within the supervised learning framework; this part is common to both classical and quantum ML \cite{bishop2006pattern, haykin2010neural}. For a dataset $\mathcal{S} = \{x_{i}, y_{i}\}_{i = 1} ^ m$ of $m$ data points where each $x_{i} \in \mathbb{R}^d$ and each $y_{i} \in \{0, 1 \}$, the classification task is to learn the optimal parameters $\theta ^ *$ of a parameterized hypothesis function $f: \mathbb{R}^d \rightarrow \{0, 1\}$ such that $\theta ^ *$ minimizes the empirical loss function defined over a subset of $\mathcal{S}$, called the training set. If $\mathcal{T} \subset \mathcal{S}$ is the training set, then the empirical loss is defined as  $ \Tilde{\mathcal{L}} = \frac{1}{|\mathcal{T}|}\sum\limits_{i = 1}^{|\mathcal{T}|} \mathcal{L}(x_{i}, y_{i})$. The optimal parameters are primarily learned using gradient-descent based methods. However, this comes with the constraint that the resulting hypothesis function $f(\theta, x)$ needs to generalize well over unseen data (the validation set $\mathcal{S} - \mathcal{T}$  as well as fresh data outside $\mathcal{S}$). This means that the learned hypothesis function should not perform substantially worse on unseen data compared to the data it has been trained on ($\mathcal{S}$), i.e., validation/ test accuracy should not be significantly lower than train accuracy. 

\subsection{Variational Classifiers}\label{var_classifiers}

\begin{figure*}[ht]
    \includegraphics[width = 2\columnwidth]{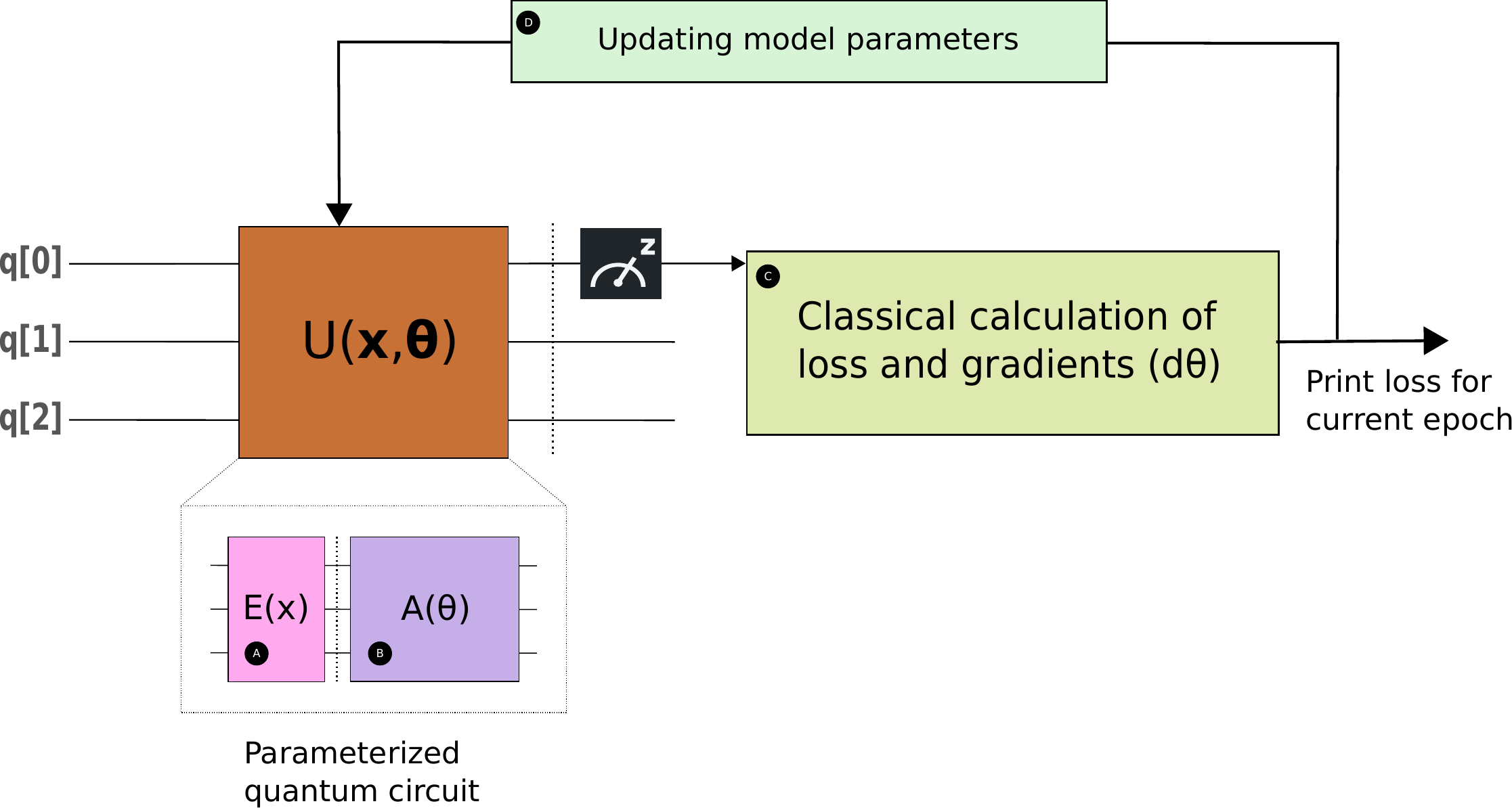}
    \caption{This figure shows the general variational circuit used for classification problems. Block A is the sub-circuit used to encode classical data into a quantum state. The resulting quantum state is acted upon by Block B, which is a set of parameterized gates. After this, we measure the quantum state and send the distribution to Block C, where the sample's loss contribution is calculated classically. This block also computes the gradients by differentiating the loss function w.r.t. the model parameters. These gradients are then used to update the model parameters, using a feedback loop in Block D.}
    \label{fig:ansatz}
\end{figure*}

A popular quantum machine learning algorithm is a hybrid quantum-classical algorithm called a variational algorithm (already introduced in Section \ref{sec:introduction}). In this sub-section, we describe a generalized variational quantum algorithm for binary classification on a classical data set \cite{schuld2020circuit, farhi2014quantum, peruzzo2014variational,markovic2020quantum,schuld2019transferlearning}. Such an algorithm uses a quantum circuit, which has a fixed ansatz (skeleton) with parameterized gates (Fig. ~\ref{fig:ansatz}). We can represent this ansatz as a parameterized unitary matrix: $U(\theta, \textbf{x})$, where $\theta$ represents the model parameter vector and $\textbf{x}$ represents the input data vector. As shown in Fig. ~\ref{fig:ansatz}, each classical data sample, taken from the training set $\mathcal{T}$, is encoded into a quantum state using schemes like amplitude encoding or qubit encoding (block A) \cite{schuld2019quantum, schuld2020circuit, havlivcek2019, schuld2017implementing}. This encoding step is followed by a parameterized quantum circuit acting on this quantum state (block B). Thus the whole ansatz $U(\theta, \textbf{x})$ can be written as a product of two matrices, i.e. $A(\theta)E(\textbf{x})$, where $E(\textbf{x})$ is the encoding sub-circuit (block A) and $A(\theta)$ is the parameterized ansatz (block B). 

Thus, analogous to the forward pass in classical neural networks \cite{lecun2015,bishop2006pattern,haykin2010neural}, an input sample in the training set $\mathcal{T}$ is passed through the quantum circuit ($U(\theta, \textbf{x})$). Then, after the measurement of the quantum state, the loss contributed by the data point to the total loss $ \Tilde{\mathcal{L}}$ is calculated classically (block C). 
Then block A, B, and C are repeated for all samples in $\mathcal{T}$ to calculate the total loss  $ \Tilde{\mathcal{L}}$. Then, in block D, the total loss is calculated and is differentiated w.r.t. the model parameters. These parameters are updated using a feedback loop by subtracting the derivatives from the original parameter values. The aim is to update the parameters such that the loss function reaches its global minima. This is done only a few times, or sometimes even once, for the entire training set $\mathcal{T}$ (See Fig. ~\ref{fig:ansatz}).  

In the above-described method, true for most variational algorithms, each sample in the classical data set is first embedded into a quantum state and then processed sequentially to calculate each sample's contribution to the loss — blocks A, B, and C. Thus, as mentioned in Section ~\ref{sec:introduction}, the time complexity (in terms of the number of samples $N$) of the algorithm, and thus the supervised learning process of the variational quantum circuit, is dominated by the steps in blocks A, B, and C, and not in block D which can be carried out for the entire batch at a time. So the time complexity of a generalized variational algorithm is given by $\mathcal{O}(N)$. All variational algorithms proposed in \cite{havlivcek2019, Tachino, schuld2017implementing, schuld2020circuit} have $\mathcal{O}(N)$ computational complexity.

\subsection{qRAM}\label{subsection: qram}
The algorithm proposed in the paper uses a qRAM, which is a memory system composed of memory cells storing quantum states. For a qRAM with $N$ different memory cells, the computational complexity for the procurement of any arbitrary superposition of data in a memory cell correlated with its address is $\mathcal{O}(\log N)$  as opposed to a classical RAM where the complexity is $\mathcal{O}(N)$ \cite{giovannetti2008quantum}. The qRAM, designed in \cite{giovannetti2008quantum}, takes as input an address register, which contains a superposition of addresses that we wish to procure $\sum\limits_{j}a_{j}\ket{j}$ and returns a superposition of data registers, correlated with the input address registers $\sum\limits_{j}a_{j}\ket{\psi_{j}}\ket{j}$. Here, $\ket{\psi_{j}}$ is a quantum state that encodes a classical data vector and is stored at the address location $j$ in the qRAM. Hence a qRAM results in the following operation:
\begin{equation}
\sum\limits_{j}a_{j}\ket{j} \xrightarrow{qRAM} \sum\limits_{j}a_{j}\ket{\psi_{j}}\ket{j}
\label{eq:qram_retrieval}
\end{equation}

qRAM is especially useful for such a task, where the data encoding scheme doesn't have any trainable parameters. Hence the resultant encoded quantum state corresponding to each sample remains fixed throughout the training process. Thus, one can encode classical data into quantum states and store them into a qRAM to not execute the encoding step for every epoch. The time complexity for retrieving a superposition of these samples correlated with their addresses as given in Eq. ~\ref{eq:qram_retrieval} takes place in $\mathcal{O}(\log N)$ time. Since we can potentially retrieve all samples in the dataset together for computation of the forward pass, the total time for the forward pass for each epoch becomes $\mathcal{O}(\log N)$.

Note that, for a standard variational algorithm described in ~\ref{var_classifiers}, storing quantum states in a qRAM would also eliminate the need to carry out the encoding step for every epoch. However, despite that, we would not get an advantage in terms of computational cost. The time complexity for retrieving any arbitrary superposition of stored quantum states from a qRAM is $\mathcal{O}(\log N)$ if $N$ quantum states are stored in it. However, for the usual classifier, the superposition would contain just one term, i.e $\ket{j} \xrightarrow{qRAM} \ket{\psi_{j}}\ket{j}$ since samples are processed sequentially. This process would have to repeat $N$ time for each epoch leading to total time complexity of $\mathcal{O}(N \log N)$. 
\section{Our proposed algorithm} \label{sec:algorithm}

In sub-section ~\ref{var_classifiers}, we had described the various "blocks" - A, B, C and D of a general variational algorithm. In this section, we describe how we have modified the three blocks A, B, and C in the above-described generalized variational algorithm to reduce the time complexity from $\mathcal{O}(N)$ to $\mathcal{O}(\log N)$.

However, before describing the four blocks, we explain how we encode classical data into quantum states and store them into a qRAM. We use an encoding with non-trainable parameters, which results in the quantum state corresponding to each classical data vector remaining fixed throughout the training process.

\vspace*{-10pt}
\subsection*{Encoding input samples in quantum states}\label{subsection: encode}

 We take a $k$-qubit quantum register initialized to the all zeros state ($\ket{0}^{\otimes k}$) and encode the vector $\textbf{x}_i$ for $ \textbf{x}_i \in \mathcal{T}$ (~\ref{cl_task}) by applying an encoding operation $E(\textbf{x})$, such that $\ket{0}^{\otimes k} \xrightarrow{E(\textbf{x}_i)} \ket{\psi_i}$ ( $\ket{\psi_i}$ is the resultant encoded state). In general $E(\textbf{x})$, may be any non-parameterized encoding scheme like amplitude encoding or qubit encoding and is carried out for all the vectors in $\mathcal{T}$. The number $k$, which is the dimension of the Hilbert space of the quantum register will vary according to the encoding scheme used. If we use amplitude encoding, then for $\textbf{x}_{i} \epsilon \mathbb{R}^d, k = \log_2 d$. If the encoding scheme is qubit encoding then $k = d$ \cite{schuld2019quantum}. 
 
 For the experiments performed in this paper, we use amplitude encoding. The resultant quantum states are then stored in a qRAM.  These stored quantum states can be retrieved at the beginning of every epoch from the qRAM, thus doing away with the need to carry out each epoch's encoding step. Thus, the process of encoding classical data to a quantum state and storing it in the qRAM becomes a one-time overhead and does not contribute to the time complexity of training the variational circuit, which accounts for costs encountered in each training epoch.
 
Now the main steps of the algorithm that are implemented repeatedly every epoch are given below. 
\subsection{\texorpdfstring{Fetching quantum states from qRAM \\(Block A)}{Fetching quantum states from qRAM (Block A)}}

The algorithm starts with a $n+k$ qubit register - the first $k$ of these qubits are called 'data qubits,' and the next $n$ qubits are called control qubits. Here $k$ is equal to the number of qubits used to encode each data vector in the encoding step. The control qubits determine the number of samples that will be used to train the parameterized ansatz parallelly. In general, for $n$ control qubits, $2^{n}$ samples can be encoded in the quantum state, half of which belong to class $0$ and the other half to class $1$. For $|\mathcal{T}| = N$, $n = \log_2 N$. This quantum state is passed through the parameterized ansatz. 

At the beginning of every epoch, to access the data samples stored in the qRAM, a layer of Hadamard gates is applied on all control qubits to prepare a superposition of contiguous qRAM addresses as shown in equation ~\ref{equation:trp_step2}. 

\begin{equation}
    \begin{aligned}
        \ket{0}^{\otimes k+n} \xrightarrow{I^{\otimes k} \otimes H^{\otimes n}} \frac{1}{\sqrt{2^{n}}}\ket{0}^{\otimes k}\Big(\sum\limits _{i = 0}^{2^{n} - 1} \ket{i}\Big)\\
    \end{aligned}
    \label{equation:trp_step2}
\end{equation}

The resulting state is passed through the qRAM to obtain a superposition of the data correlated with the addresses. If $N$ entries are stored in the qRAM, this step takes $\mathcal{O}(\log N)$ steps; while a classical RAM takes $\mathcal{O}(N)$ steps for the same \cite{giovannetti2008quantum}.

\begin{equation}
    \begin{aligned}
        \frac{1}{\sqrt{2^{n}}}\ket{0}^{\otimes k}\Big(\sum\limits_{i = 0}^{2^{n} - 1} \ket{i}\Big) \xrightarrow{qRAM} \frac{1}{\sqrt{2^{n}}}\Big(\sum\limits_{i = 0}^{2^{n} - 1}\ket{\psi_{i}} \ket{i}\Big)
    \end{aligned}
    \label{equation:trp_step3}
\end{equation}

\subsection{\texorpdfstring{Operation of the parameterized quantum circuit (Block B)}{Operation of the parameterized quantum circuit (Block B)}}\label{subsection: param_op}

After this the parameterized circuit $A(\theta)$ acts on the data qubits to obtain the final desired state. The operation $A(\theta)$ acts on a state such that $\ket{\psi_{i}}\xrightarrow{A(\theta)}\ket{\Psi_{i}}$

\begin{equation}
    \begin{aligned}
        \frac{1}{\sqrt{2^{n}}}\Big(\sum\limits_{i = 0}^{2^{n} - 1}\ket{\psi_{i}}\ket{i}\Big)\xrightarrow{A(\theta) \otimes I^{\otimes n}}
        \frac{1}{\sqrt{2^{n}}}\Big(\sum\limits_{i = 0}^{2^{n} - 1}\ket{\Psi_{i}}\ket{i}\Big)
    \end{aligned}
    \label{equation:trp_step4}
\end{equation}

Here, the state $\ket{\Psi} = \frac{1}{\sqrt{2^{n}}}\Big(\sum\limits_{i = 0}^{2^{n} - 1}\ket{\Psi_{i}}\ket{i}\Big)$ is called the \textit{Data state}.

As in any variational quantum circuit we wish to  to tune the model parameters, $\theta$ such that if any $x_{i}$, has $y_{i}$ =0 (class 0) then the $\ket{\psi_{i}}$ tends to $\ket{0}$. Similarly, for any $x_{i}$, if $y_{i}$ =1 (class 1) then the $\ket{\psi_{i}}$ tends to $\ket{1}$.

\begin{figure}[ht]
    \includegraphics[width = 1.1\columnwidth, left]{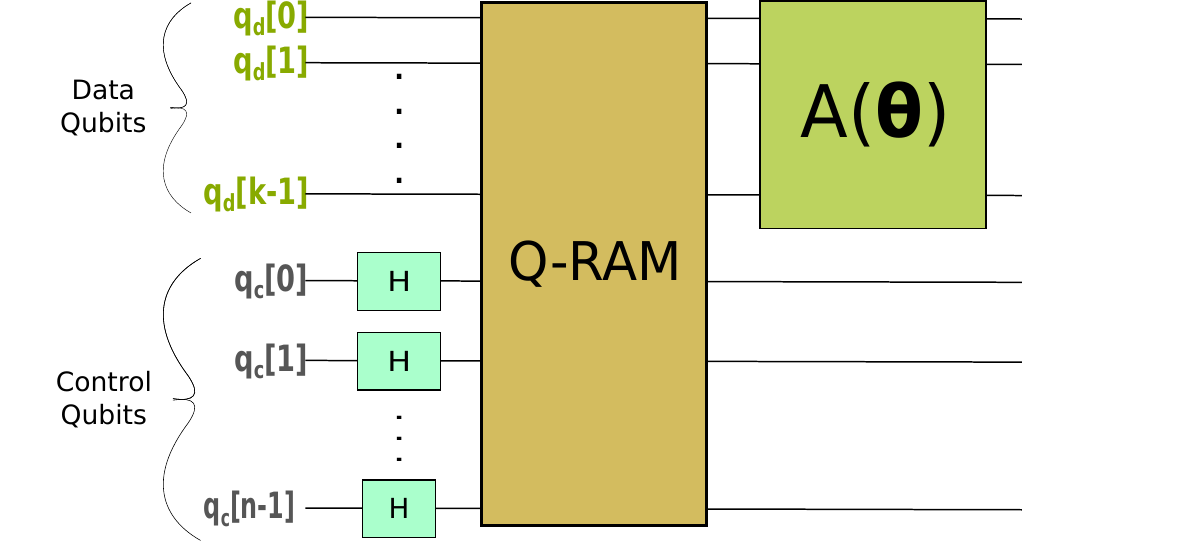}
    \caption{This figure shows the circuit used to prepare the Data state. The first component is the qRAM, which is used to retrieve multiple samples correlated with their respective addresses as a superposition state. This resulting state is acted upon by a set of parameterized gates-$A(\theta)$ to give the final Data state.}
    \label{fig:training_state_prep_circuit}
\end{figure}

\subsection{Loss calculation (Block C)}\label{subsection: lc}
\subsubsection{Label State Preparation}\label{subsubsection: label_prep}

We first define a new quantum state corresponding to the data state which we obtained at the end of ~\ref{subsection: param_op}. We call it the \textit{Label state}. The label state is a $n+1$ qubit state with 1 data qubit and $n$ control qubits ($n = \log_2 N$). As the name suggests, the label state stores the value of the label corresponding to a given data-point and is required for embedding loss calculation in the quantum circuit itself. 
For the Label State preparation, we start with a $n+1$ qubit quantum register, one of which is a data qubit and the rest $n$ are control qubits, initialized to $\ket{0}^{\otimes n+1}$.
We apply a layer of Hadamard gates to the $n$ control qubits. 

\begin{equation}
    \ket{0}^{\otimes n+1} \xrightarrow{(I \otimes H^{\otimes n})} \frac{1}{\sqrt{2^{n}}}\ket{0}\Big(\sum\limits_{i = 0}^{2^{n} - 1} \ket{i}\Big)
    \label{equation:tsp_step1}
\end{equation}

This state can also be written as:
\begin{equation}
    \frac{1}{\sqrt{2^{n}}}\ket{0}\Big(\sum\limits_{i_{0} = 0}^{1}\sum\limits_{i_{1} = 0}^{1} ....\sum\limits_{i_{n-1} = 0}^{1}\ket{i_{0}}\ket{i_{1}}...\ket{i_{n-1}}\Big)
    \label{equation:tsp_step2}
\end{equation}

Now we apply a CNOT gate from control qubit $0$ to the data qubit.

\begin{equation}
    \begin{aligned}
         & \frac{1}{\sqrt{2^{n}}}\ket{0}\Big(\sum\limits_{i_{0} = 0}^{1} ....\sum\limits_{i_{n-1} = 0}^{1}\ket{i_{0}}...\ket{i_{n-1}}\Big)\\
         & \xrightarrow{CNOT(\ket{i_0}), \ket{0})} \frac{1}{\sqrt{2^{n}}}\Big(\sum\limits_{i = 0}^{2^{n-1} - 1}\ket{0}\ket{i} + \sum\limits_{i = 2^{n-1}}^{2^{n} -    1}\ket{1}\ket{i}\Big)
    \end{aligned}
    \label{equation:tsp_step3}
\end{equation}

This final state is the label state, and we denote it by $\ket{\Phi}$. For half of the $2^{n}$ possible values that the control qubit register can take, the data qubit value is $0$, and for the other half, it is $1$. This is because half the samples encoded in the data state have a label of $0$, and the other half have a label of $1$. Our goal is to change the model parameters $\theta$ in such a way that the data state "approaches" the label state. This can be achieved if the inner product of these two states is equal to 1 up to an overall phase, i.e. $\abs{\bra{\Psi}\ket{\Phi}} \xrightarrow{} 1$.  

\begin{figure}[ht]
    \includegraphics[width = \columnwidth, left]{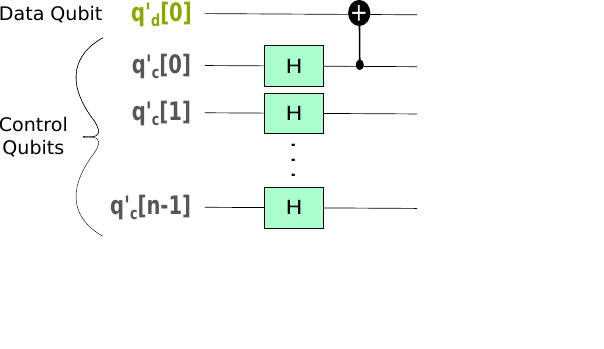}
    \caption{This figure shows the circuit used to prepare the Label state.}
    \label{fig:target_state_prep_circuit}
\end{figure}

\subsubsection{Embedding Loss Calculation in the Quantum Circuit}\label{subsubsection:embedding loss}
We discussed how we wanted $\abs{\bra{\Psi}\ket{\Phi}} \xrightarrow{} 1$. This can be achieved by minimizing the following loss.

\begin{equation}
    \mathcal{L} = 1 - \abs{\bra{\Psi}\ket{\Phi}}^{2}
    \label{eq:loss_calculation}
\end{equation}

The maximum loss that can be incurred is 1 (when the two states are orthogonal), and the minimum is 0 (for the case when the two states are the same up to a global phase). This loss can be implemented efficiently using a multi-qubit swap test circuit \cite{garcia2013swap}. While we find a rigorous analysis of the swap-test-based loss function as out of the scope of this paper, the idea of embedding the loss calculation in the quantum circuit itself rather than performing a measurement on the output of a quantum variational circuit and calculating the loss classically has been explored in \cite{cao2020cost}.  

Further, the computational complexity for the swap test circuit rises linearly with the number of qubits in the circuit (as can be inferred from  \ref{fig:swap_test_circuit}). The number of qubits is $\mathcal{O}(\log N)$. Thus, the swap test circuit's computational complexity for implementing this loss is also $\mathcal{O}(\log N)$ for a dataset with $N$ samples. The Swap test is a standard quantum computing circuit which can be used to quantify the "closeness" between two quantum states. It also has a computational cost of $\mathcal{O}(\log N)$, all of which make it a suitable choice as a loss calculation circuit.

Hence our algorithm exponential advantage over any classical ML model where loss calculation takes $\mathcal{O}(N)$ time for $N$ samples. Together with the qRAM, which takes $\mathcal{O}(\log N)$ time to retrieve $N$ data points and the application of the parameterized unitary which takes $\mathcal{O}(1)$ time for $N$ samples, the overall computational complexity of the forward pass becomes $\mathcal{O}(\log N)$ which is an exponential advantage over any classical ML model.

\subsection{Loss optimization (block D)}\label{subsection: wu}

After calculating loss for all data points, we obtain the gradient of this loss w.r.t. the classifier parameters $(\theta)$. For the experiments done in this paper, we calculate these gradients numerically. Using these gradients, we update all model parameters using gradient descent. This process is repeated once per epoch till the loss converges. Hence, this process does not determine the complexity of the algorithm, as explained earlier.

\begin{figure*}[t!]
    \includegraphics[width = 2\columnwidth]{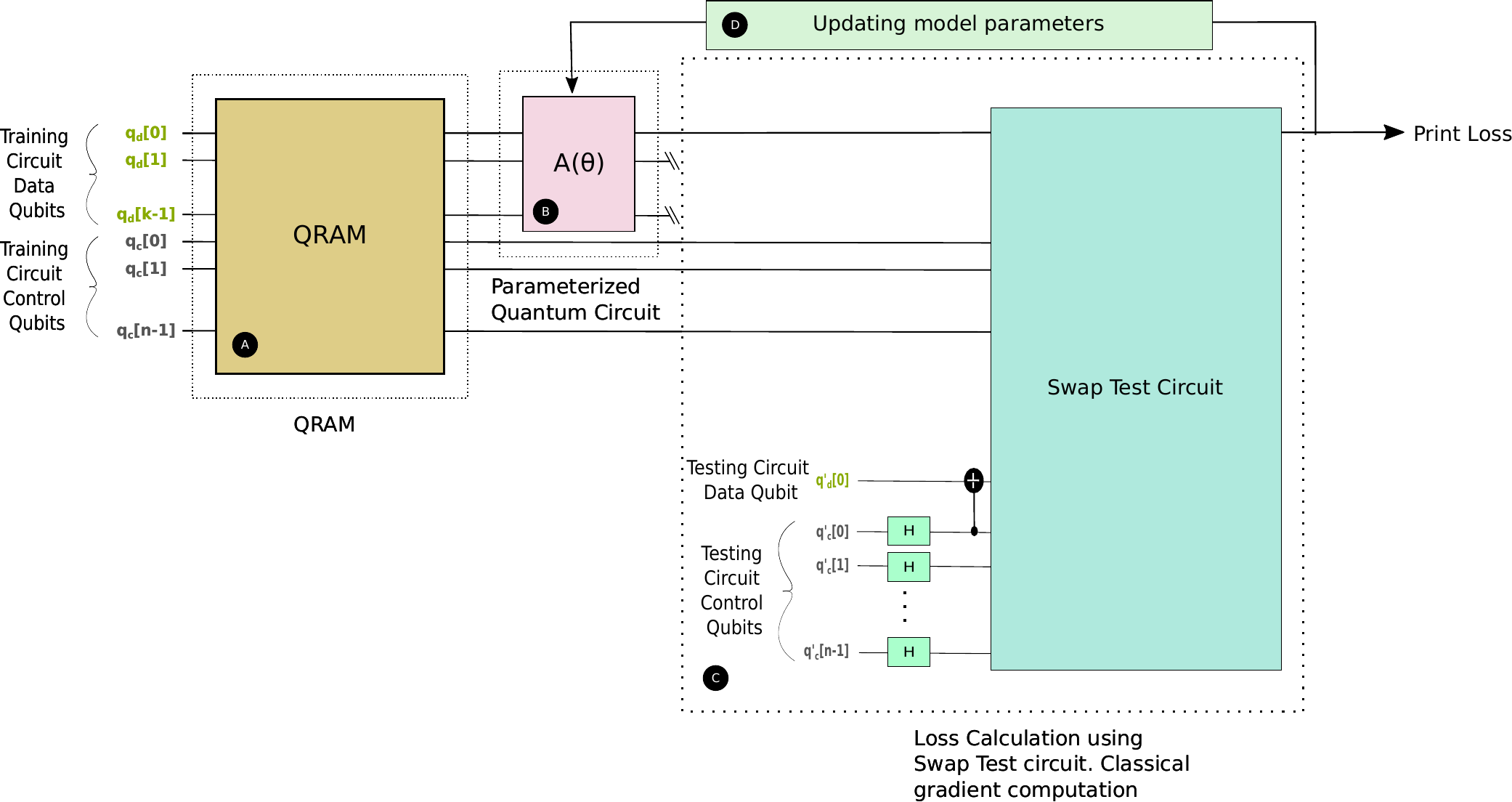}
    \caption{This figure shows the full circuit implementing the proposed algorithm. Block A is the qRAM, which is used to retrieve multiple encoded samples simultaneously. Block B is a set of parameterized gates that acts on the quantum states encoding the data. Block C is used to calculate the loss using quantum circuits. The gradients are computed by differentiating the loss classically. The model parameters are updated in Block D using a feedback loop. In terms of the individual blocks' functionality, we can draw an analogy between the blocks in this figure and Fig. ~\ref{fig:ansatz}.}
    \label{fig:full_circuit}
\end{figure*}

\vspace{5mm}
\begin{figure}[ht]
    \includegraphics[width = \columnwidth, left]{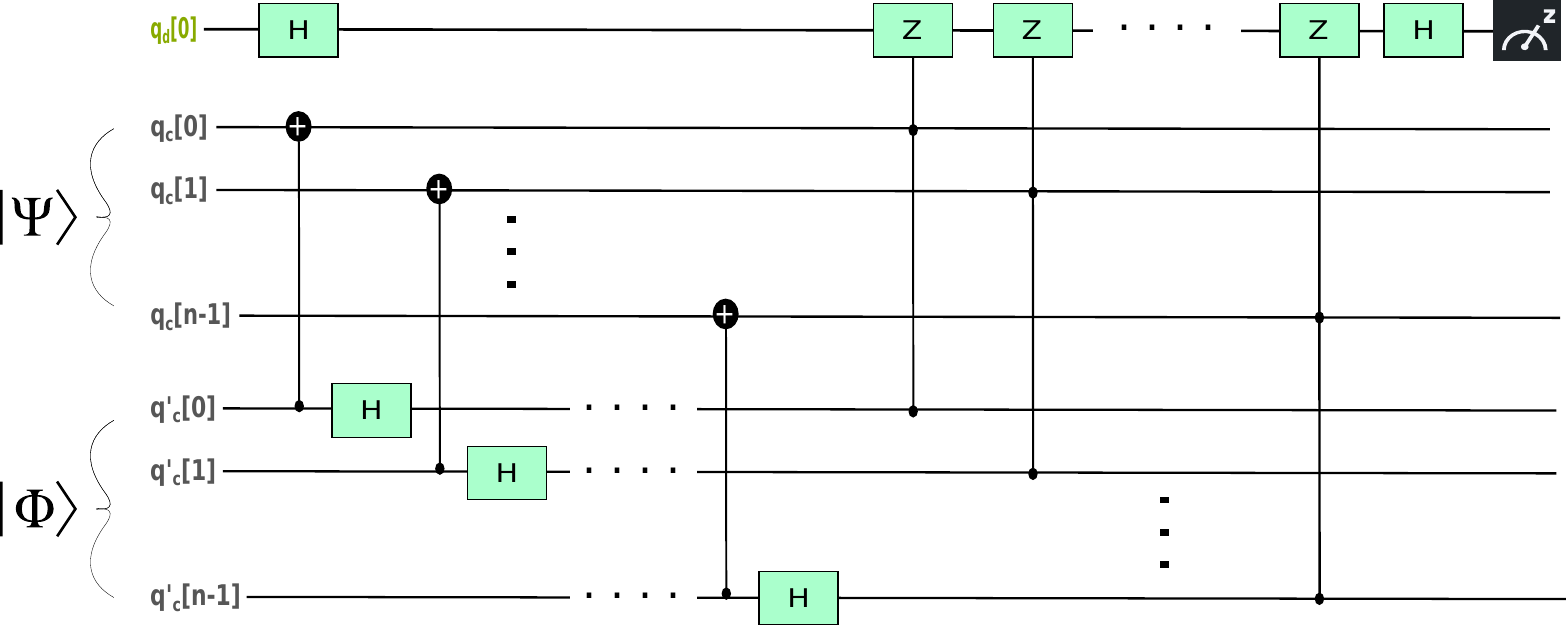}
    \caption{Swap Test Circuit used for calculating the absolute value of the inner product of $n$ qubit Data State and Label State. Here, $\ket{\Psi}$ denotes the data state while $\ket{\Phi}$ denotes the label state.}
    \label{fig:swap_test_circuit}
\end{figure}

We have designed quantum circuits to implement the different blocks-A, B, and C (D is a classical step). Fig. ~\ref{fig:training_state_prep_circuit} shows Block A and Block B. The retrieval of the quantum state containing the training data points is done using the qRAM, which is Block A. The parameterized ansatz $A(\theta)$ is Block B. Fig. ~\ref{fig:target_state_prep_circuit} shows the label state preparation circuit. This circuit implements the steps of the algorithm described in ~\ref{subsubsection: label_prep}. For calculation of the absolute value of the inner product, described in ~\ref{subsubsection:embedding loss} we use the swap test circuit as shown in Fig. ~\ref{fig:swap_test_circuit}. This loss is then differentiated, and the derivatives are used to update the parameters using a classical feedback loop. The entire circuit schematic can be seen in Fig. ~\ref{fig:full_circuit}.
\section{Numerical Experiments}\label{sec:experiments}

We evaluate the proposed algorithm for binary classification by analyzing the model performance on Fisher's Iris dataset \cite{fisher1936use}. Fisher Iris is a 3 class data set (classes-Setosa, Virginica, and Versicolor) consisting of 150 data points, 50 belonging to each class. Out of the three classes, classes Virginica and Versicolor are linearly inseparable w.r.t. each other while class Setosa is linearly separable w.r.t both of them. Each data point is a four-dimensional vector. We perform binary classification for all three possibilities (class Setos vs. class Virginica, class Virginica vs. class Versicolor, class Setosa vs. class Versicolor). Thus each classification task uses 100 data points, out of which 80 are used for training the circuit while 20 are used for testing the performance of the trained model on unseen data. We implement the functionality of the proposed quantum algorithm on a classical computer by performing simulations in Python. For the experiments, we use $n = 2$; this means that four data samples, two of each class, are fed into the circuit "simultaneously" to train it. The scheme used in our experiments to encode the classical data vectors into quantum states is amplitude encoding. For a normalized vector, $(x_1, x_2, x_3, x_4) \xrightarrow{Amplitude Encoding} \sum_{i = 1}^{4}x_i\ket{i}$

The results for all possible binary classification for the case of Fisher Iris dataset is given below  
\begin{center}
\begin{tabular}{ |c|c|c|c| } 
 \hline
 Class 0 & Class 1 & Training Acc. & Testing Acc.\\
 \hline
 Setosa & Versicolor & 1.0 & 1.0 \\ 
 Virginica & Versicolor & 0.925 & 0.95 \\
 Setosa & Virginica & 1.0 & 0.95 \\ 
 \hline
\end{tabular}
\end{center}

\begin{figure}[ht]
    \includegraphics[width = 1.02\columnwidth, left]{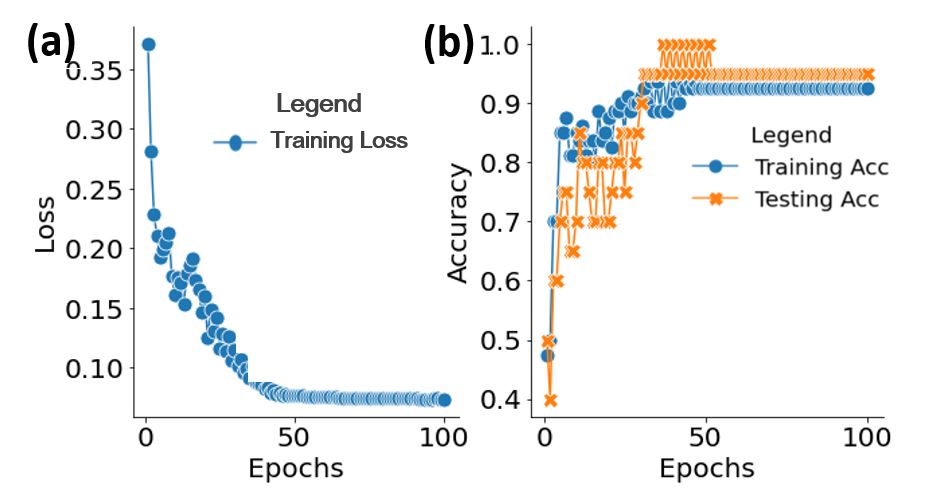}
    \caption{(a) Training loss for Fisher Iris in case of Virgenica vs Versicolor across Epochs (b) Training and Testing Accuracy for Fisher Iris in case of Virgenica vs Versicolor across Epochs}
    \label{fig:Training_loss_FisherIris1}
\end{figure}

Fig. ~\ref{fig:Training_loss_FisherIris1} shows the training loss and training and testing accuracy for classification between Versicolor and Virginica.

Overall, we evaluated our model classifier on three different binary classification problems using the multi-sample training scheme presented in this paper. We have reported the results of the classification of samples from a popular data-set among the QML community, Fisher's Iris, and observe fairly high training and testing accuracy, which is in line with the results obtained with training a generic Variational Quantum Circuit for the same. We have plotted the results of training the given variational quantum circuit with four samples at a time for Training Loss, Training Accuracy, and Testing Accuracy of the model across 100 epochs.
\section{Discussion: Strengths and limitations of the algorithm}\label{sec:discussion}

Following are the benefits of the algorithm over classical machine learning and contemporary variational algorithms and its primary contributions to quantum machine learning.

\begin{itemize}
    \item \textbf{Reduction in computational cost of loss calculation step}: As mentioned earlier, the computational cost of the feed-forward step for $N$ samples is $\mathcal{O}(\log N)$ as opposed to the classical case where the computational cost is $\mathcal{O}(N)$. 
    \item \textbf{Agnosticism to the classifier ansatz}: The algorithm is agnostic to the classifier ansatz used $(A (\theta))$. Any parameterized classifier can be used as Block B in the overall circuit schematic (Fig. ~\ref{fig:full_circuit}). The only constraint is that the encoding of classical data to a quantum state should not involve any trainable parameters. One thus has the flexibility to use any parameterized ansatz of choice.
\end{itemize}

The following are certain limitations and open questions about the algorithm that can be worked on to improve it.

\begin{itemize}
    \item \textbf{Training on a non-standard loss}: The loss function calculated using the circuit imposes a penalty if a data sample with target class '1' is classified as '0' with a non-zero probability. The same holds for a sample with target class '1' being classified as '0' with a non-zero probability. This loss makes intuitive sense and works for the experiments reported but doesn't reduce to a standard loss function (like cross-entropy). Standard loss functions have been well studied well and have a vast body of literature supporting their use for convex optimization tasks. This rigorous analysis is lacking for the loss we use. However, it does work well for the experiments conducted on Fisher's Iris dataset.
    
    \item \textbf{Separate circuits for separate losses}: The swap test circuit which we use for calculation of loss is useful only if the loss is the one used in the paper, i.e., $1 - \abs{\bra{\Psi}\ket{\Phi}}^{2}$. For some other loss, a different loss calculation circuit has to be designed. Further, to ensure that the entire circuit's time complexity remains the same, one needs to ensure that the new loss calculation circuit has a computational complexity of at most $\mathcal{O}(\log N)$ for a dataset of $N$ samples. 
\end
{itemize}
\section{Conclusion}\label{sec:conclusion}
In conclusion, we have proposed a new algorithm that can train any arbitrary quantum variational classifier in $\mathcal{O}(\log N)$ time for a dataset of $N$ samples, as opposed to a classical ML algorithm or standard variational algorithm that take $\mathcal{O}(N)$ time for the same. This reduction in training complexity is achieved primarily because of the complexity reduction that is achieved by the use of qRAM that can retrieve $N$ samples in $\mathcal{O}(\log N)$ operations and the usage of the swap-test circuit for calculating the loss, a procedure that again takes place in $\mathcal{O}(\log N)$ time. Together, with other $\mathcal{O}(1)$ time operations, we can carry out the "forward pass" in $\mathcal{O}(\log N)$ time. This speedup is orthogonal to the potential speedup in processing an individual data vector that a variational classifier can potentially achieve. The speedup that we obtain is because of the processing of multiple samples simultaneously during training. This ability to process multiple samples (potentially the entire dataset) simultaneously is useful in processing enormous machine learning datasets with millions of data points. Ref. ~\cite{adhikary2020entanglement} proposes a similar training algorithm that processes two samples simultaneously to speed up training.

Although the discussion in this paper covers only binary classification, the algorithm can very easily be generalized for multi-class classification problems by increasing the data qubits in the label state preparation circuit (For two qubits, for example, we can solve a 4 class classification problem with '00', '01', '10', '11' as the four labels). The algorithm uses a non-standard loss function for the optimization task. An open problem is to design a circuit with at most $\mathcal{O}(\log N)$ complexity for calculating the loss of $N$ data points that also translates into a standard loss for classification task (like cross-entropy loss).

\bibliography{sections/References}
\end{document}